\documentclass[two column, bibtex]{revtex4-2}
\usepackage[utf8]{inputenc}

\usepackage{mathtools}
\usepackage{amssymb}
\usepackage{graphicx}
\usepackage{xspace}
\usepackage{hyperref}

\begin{document}

\newcommand{\ie}{\textit{i.e.}\xspace}

\title{Effective-range expansion of the $T_{cc}^+$ state at the complex $D^{*+}D^0$ threshold}
\date{March 2022}


\author{Mikhail Mikhasenko}
\affiliation{ORIGINS Excellence Cluster, 85748 Garching, Germany}

\begin{abstract}
Evaluation of the effective-range parameters for the $T_{cc}^+$ state in the LHCb model is examined.
The finite width of $D^*$ leads to a shift of the expansion point into the complex plane to match analytical properties of the expanded amplitude.
We perform an analytic continuation of the three-body scattering amplitude to the complex plane in a vicinity of the branch point and
develop a robust procedure for computation of the expansion coefficients.
The results yield a nearly-real scattering length, and two contributions to the the effective range which have not been accounted before.
\end{abstract}

\maketitle

\section{Introduction}

A low-energy behavior of the quantum chromodynamics remains one of the most intriguing research area of the particle physics.
The physics at this regime is governed by the effective degrees for freedom, massive constituent quarks forming
hadronic states. Significant progress in understanding properties of the hadronic interaction has been achieved in the last 70 years
of the scrupulous research~\cite{ParticleDataGroup:2020ssz}. The recent rise of the interest, however, is caused
by a series of discoveries of new types of hadrons beyond the conventional mesons and baryons~\cite{Olsen:2017bmm,Brambilla:2019esw}.
In the last 20 years, the hadronic molecules~\cite{Guo:2017jvc} and multiquark states~\cite{Esposito:2016noz,Ali:2019roi} moved from the theoretical concept to genuinely existing states abundantly produced in the modern particle-physics experiments, see also the reviews~\cite{Chen:2016qju,Ali:2017jda,Liu:2019zoy}.
The field of hadron spectroscopy is actively driven by an intense rate of discoveries of new exotic states
starting from $\chi_{c1}(3872)$ in 2003~\cite{Belle:2003nnu}, and going through many observations of $X$,$Y$,$Z$, and $P_c$ families, as well as, discovery of the $T_{cc}^+$ state in the middle of 2021~\cite{LHCb:2021auc,LHCb:2021vvq}. Heated scientific debates are held around the properties and microscopic nature of these states.

The $T_{cc}^+$ tetraquark observed by the LHCb experiment~\cite{LHCb:2021vvq,LHCb:2021auc} 
in decays to $D^0D^0\pi^+$ final state has a minimal quark content of $(cc\bar{u}\bar{d})$. The mass is found to be around $360\,$keV below the $D^{*+}D^0$ threshold, and the width of several dozens of keV. The state was greatly anticipated based on quark-model calculations~\cite{
SilvestreBrac:1993ss,%
Pepin:1996id,%
Lee:2009rt,%
Janc:2004qn,%
Ebert:2007rn,%
Vijande:2003ki,%
Luo:2017eub,%
Park:2018wjk,%
Deng:2018kly,%
Yang:2019itm,%
Maiani:2019lpu,%
Tan:2020ldi,%
Lu:2020rog,%
Faustov:2021hjs,%
Noh:2021lqs},
using Bethe-Salpeter equation~\cite{Feng:2013kea},
QCD sum rules~\cite{Navarra:2007yw,Wang:2017uld,Gao:2020ogo},
diquark modelling~\cite{Semay:1994ht,Carlson:1987hh},
molecular considerations~\cite{Li:2012ss,Liu:2019stu},
as well as in phenomenological approaches~\cite{Cheng:2020wxa,Karliner:2017qjm,Braaten:2020nwp}
While many models support the existence of $T_{cc}^+$, they conflict in their comprehension of the state formation.
There is no consensus either $T_{cc}^+$ is a genuine QCD state similar to other mesons and baryons strongly coupled to continuum,
or it is a purely molecule state formed by $D^*$ and $D$ bound together by the residual color-neural strong forces,
similar to deuteron or nuclear atoms.

The Weinberg's compositeness criterion, an approach sensitive to the internal structure of the states,
was adopted to hadron states by the authors of~\cite{Baru:2003qq}.
According to Weinberg's consideration~\cite{Weinberg:1962hj,*Weinberg:1965zz},
the low-energy scattering parameters, the scattering length and effective range~\cite{Bethe:1949yr} reflect the microscopic inner working
of the near-threshold hadrons.
Namely, the non-relativistic scattering theory establish a general form of the two-particle scattering amplitude, $T(s)$,
\begin{align} \label{eq:scatt.ampl}
    T(s) = \frac{N}{R(s) - ik(s)}\,,
\end{align}
where $m_a$ and $m_b$ are the particle masses, $s$ is the squared mass of the system,
$k$ is a break-up momentum, $k = \lambda^{1/2}(s,m_a^2,m_b^2)/2\sqrt{s}$,
with $\lambda$ being the K{\"{a}}ll{\'{e}}n function~\cite{Kallen},
and $R(s)$ is a reaction-specific energy dependence of the scattering amplitude.
The break-momentum is zero when the squared mass of the system is equal to $\text{thr}=(m_a+m_b)^2$. 
The effective-range expansion is introduced to describe the behavior of the amplitude near the threshold, \ie for small values of the momentum $k$.
The Taylor series of the function $R$ reads:
\begin{align} \label{eq:R.NR}
    R(s) = \frac{1}{a} + r\frac{k^2(s)}{2} + O(r^2k^4)\,.
\end{align}
where $a$ is the scattering length, and $r$ is the effective range of the particle interaction.
For the sake of the following discussion, we consider the case of near threshold state with $a<0$.
Analytic structure of $R$ implies no odd powers of $k$ in the series
since $k(s)$ has a branch point singularity at $s=s_\text{thr}$, while $R(s)$ is regular at this point.
The compositeness $X$ is computed as
\begin{align} \label{eq:compositeness}
    X = 1/\sqrt{1+2r/a}\,,
\end{align}
up to corrections due to the interaction range.
Small compositeness indicates a genuine compact QCD state, however the limit $X\to 0$ in unphysical as the state decouples from the continuum (becomes unobservable). On the other end, $X = 1$ corresponds to a pure molecule~\cite{Weinberg:1965zz}.
Moreover, for potential scattering, the effective range is strictly positive~\cite{Smorodinsky:1948xyz, *Landau:1991wop},
which leads to the values of $X$ more of equal to $1$. The values $X>1$ are unphysical either~\cite{Wigner:1955zz} and appear,
e.g. for deuteron, due to unaccounted interaction range in Eq.~\eqref{eq:compositeness}.
The authors of~\cite{Esposito:2021vhu} point that the negative effective range, i.e. $X\neq 1$ indicates a elementary component of the state. However, the author of~\cite{Bruns:2019xgo} interprets the high value of $X$ as a probability to find the molecular constituents separated by a distance greater than the interaction range which does not have to be identically one for a pure molecule.
Computation of the compositeness for many practical cases, e.g. in presence of the higher thresholds is currently under debates~\cite{Baru:2021ldu}.
Various generalization of the Weinberg compositeness has been offered, e.g.
\cite{Sekihara:2014kya} applies compositeness to unstable resonances,
\cite{Matuschek:2020gqe} considers compositeness of virtual states,
\cite{Li:2021cue} suggests an expression for compositeness using the scattering phase,
\cite{Song:2022yvz} discusses the cases with very large binding energy.

Effective-range parameters for the $D^*\bar{D}$ and $D^*D$ scattering have gained a lot of attention recently.
The LHCb analysis of the $\chi_{c1}(3872)$ lineshape~\cite{LHCb:2020xds} in vicinity of the $D^{*0}\bar{D}^0$
follows parametrization of \cite{Hanhart:2007yq}.
The inverse scattering length is measured with a large uncertainty in the imaginary part due to purely-constrained inelastic channels. A wide confidence interval for the effective range is inferred from the reported values of the coupling~\cite{Baru:2021ldu}.
The $T_{cc}^+$ analysis of LHCb~\cite{LHCb:2021vvq} offers an amplitude of high complexity where the unstable $D^*$ is considered,
there are three-body effects related to one-particle exchange, and well as singularities from the $D^{*0}D^+$ threshold.
However, the effective-range parameters are obtained in a simplified manner.
The scattering length, $a_r$ is estimated from the value of the inverse scattering amplitude on the real axis at $s=(m_{D^{*+}}+m_{D^0})^2$:
\begin{align} \label{eq:a.LHCb}
    1/a_\text{r} = (-25.8 \pm 2.1) - 6.7i\,\text{MeV}\,.
\end{align}
Here, the uncertainty is related to the fitting error of the $T_{cc}^+$ mass.
As commented in~\cite{LHCb:2021vvq} the imaginary part of $a_r$ is not related to the details of the $D^{*+}D^0$ scattering,
while it is a simple consequence of the finite width of the $D^{*+}$ meson.
The confidence interval for the effective range is computed from the limits on the coupling of the potential (K-matrix) term.
\begin{align} \label{eq:r.LHCb}
    -11.9(-16.9) < r_\text{pot.} < 0\,\text{fm}\,\,\text{at}\,\,90(95)\%\,\text{CL}\,.
\end{align}
The author of \cite{Albaladejo:2021vln} performs an analysis of the binned LHCb distributions using
the Lippmann–Schwinger setup with a potential that includes only a contact term.
The effective-range parameters are determined by
a fit of the effective-range expression to the model in vicinity of the threshold.
Both the scattering length and the effective range strongly depend on the value of the cutoff.
While the scattering length is comparable with the value of the LHCb analysis,
the effective range is strictly positive and equal to $0.63\,$fm ($1.26\,$fm) for the cutoff of $1\,$GeV\,($0.5\,$GeV).
Ref.~\cite{Du:2021zzh} perform a comparable analysis of the LHCb data with more advanced model including the one-pion exchange potential.
The value of $r$ is determined by a derivative of the inverse amplitude with respect to the system energy on the real axis in accordance with Eq.~\eqref{eq:R.NR}. For the nominal model, it is found to be $-2.40 \pm 0.85\,$fm with the uncertainty determined by the cutoff dependence.

It has been realized in the recent discussions~\cite{meeting.CERN, meeting.Munich} that
there are contributions to the effective range of the $T_{cc}^+$ state in addition to the potential term Eq.~\eqref{eq:r.LHCb} that might be important.
In what follows, we define a mathematically accurate effective-range series that approximate the amplitude in the complex plane
in a large region, up to the next inelastic threshold. The method is implemented for the $T_{cc}^+$ model of the LHCb analysis
solving list of technical issues. The effective-range parameters are presented and discussed.


The LHCb model of the $T_{cc}^+ \to D^0D^0\pi^+$ transition is derived in \cite{LHCb:2021auc}.
The construction is inspired by the model-independent approach based on three-body unitarity
and a factorization assumption~\cite{Mikhasenko:2019vhk}.
Specifically, the $D^{*+}D^0 \to D^{*+}D^0$ scattering amplitude reads:
\begin{align}\label{eq:Tcc.model}
    \mathcal{T}(s)= \frac{g^2}{m^2 - s - g^2\Sigma(s)}\,.
\end{align}
The expression is reminiscent of the relativistic Breit-Wigner parametrization with a non-trivial self-energy function, $\Sigma$,
which accounts for three coupled channels, $D^{0}D^{0}\pi^+$, $\pi^0D^{+}D^{0}$, and $\gamma D^{+}D^{0}$.
The imaginary part of the self-energy is computed as a sum of the averaged squared matrix elements of the $T_{cc}^+$ decays,
\begin{align}
\rho_\text{tot} = \rho_{\pi^+ D^0 D^0}+\rho_{\pi^0 D^+ D^0}+\rho_{\gamma D^+ D^0}\,.    
\end{align}
Once-subtracted dispersion relations are used to calculate the full loops function $\Sigma(s)$:
\begin{align}
    \Sigma(s) = \Sigma_0 + \frac{s}{2\pi i} \int_{s_\text{thr}}^\infty \frac{i\rho_\text{tot}(s')\,d s'}{s'(s'-s-i0)}\,.
\end{align}
Here, the subtraction value $\Sigma_0$ ensures that the amplitude is purely imaginary for $s=m^2$.
The first order polynomial $m^2-s$ in the denominator of Eq.~\eqref{eq:Tcc.model} originates from the K-matrix pole, $g^2/(m^2-s)$.
The integration over the three-body phase space is done numerically using the expression,
\begin{align} \label{eq:rho.i.complex}
    \rho_f(s) = \frac{4m_{23} m_{12}}{2\pi(8\pi)^2 s} \int_{\mathcal{D}_f} dm_{12} dm_{13}\,\frak{M}_f^2\,,
\end{align}
where $m_{12}$ and $m_{13}$ are invariant masses of different particle pairs for the final state labelled by $f$,
$\frak{M}_f^2$ is a squared matrix element of the $T_{cc}^+ \to f$ decay,
and $\mathcal{D}_f$ is the phase-space integration domain in $m_{12}$ and $m_{13}$ variables.
The matrix element has the form
\begin{align}
    \frak{M}_f^2 = \frak{F}_f^\dagger\,\frak{X}_f(s,m_{12}^2,m_{13}^2)\,\frak{F}_f\,,
\end{align}
where $\frak{F}_f$ is a two-element vector of $D^*$ amplitudes with the first being for the $D^*$ propagator in the $m_{12}$ variable, and the second for the $m_{13}$ variable.
The quantity $\frak{X}_f(s,m_{12}^2,m_{13}^2)$ stands a two dimensional squared matrix of polynomial in the Mandelstam variables,
$m_{12}^2$ and $m_{13}^2$ and $s$ given in Appendix~\ref{sec::Tcc.model}.


The function in Eq.~\eqref{eq:Tcc.model} has three branch points on the real axis that correspond to
the $D^{0}D^{0}\pi^+$, $\pi^0D^{+}D^{0}$, and $\gamma D^{+}D^{0}$ thresholds.
No other singularities emerge in the entire complex plane, the first Riemann sheet, due to the use of the dispersion relation.
The resonance pole is located at the second Riemann sheet below the real axis.
For computation of the second-Riemann-sheet values, the inverse amplitude $\mathcal{T}^{-1}$ is supplemented by the real-axis discontinuity of $-\Sigma(s)$ given by $i\rho_\text{tot}(s)$. The expression,
\begin{align}
    \mathcal{T}^{-1}_{I\!I}(s) = 
        \mathcal{T}^{-1}(s) + i\rho_\text{tot}(s)\,,
\end{align}
is computed for the complex values of $s$ with negative imaginary part to access the resonance region at the complex plane.

\begin{figure*}
    \centering
    \includegraphics[width=\textwidth]{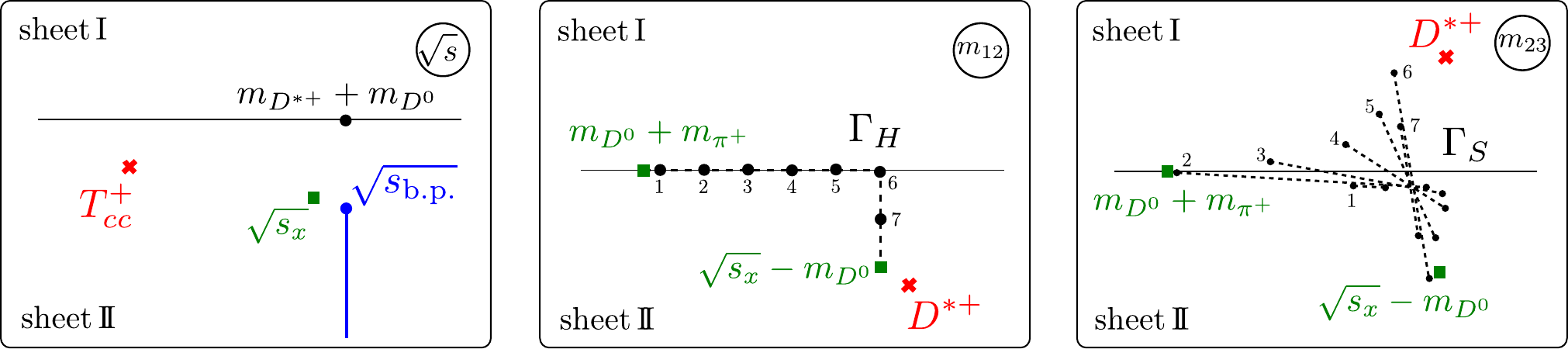}
    \caption{A schematic representation of the analytic structure and the integration paths in $\sqrt{s}$, $m_{12}$ and $m_{13}$ variables for the functions $\rho_i(s)$, $\frak{F}(m_{12})$ and $\frak{F}^{\dagger}(m_{13})$, respectively.
    The poles are shown by the red crosses, see $T_{cc}^+$ pole in the $\sqrt{s}$ plane (left) and $D^{*+}$ poles in the
    $m_{12}$ and $m_{13}$ planes
    The blue point indicates the complex $D^{*+}D^{0}$ branch point. The blue line shows the branch cut
    oriented straight down from the branch point. Dashed lines outline the integration paths for calculation of the
    the phase-space integral at the value $s = s_x$ in vicinity of the branch point $s_\text{b.p.}$.
    The integration path in the $m_{12}$ variable connects the integral end points using the ``hook'' prescription.
    The integration in the $m_{13}$ variable is preformed along a straight line which connects the $m_{12}$-dependent end points.
    These paths are shown for seven values of $m_{12}$ labelled by the numbers from $1$ to $7$.
    }
    \label{fig:cuts}
\end{figure*}

The $D^{*+}D^0$ branch point denoted by $s_\text{b.p.}$.
It emerges from the $\rho_\text{tot}$ function when the integration domain is hit the pole of $D^{*+}$ (see Fig.~\ref{fig:cuts}).
The location of this singularity in $\sqrt{s}$ variable is given by a coincidence of the upper limit of the phase-space range
$m^{(\text{max})} = \sqrt{s}-m_{D^0}$ with
the $D^{*+}$ pole in the $\frak{F}(m)$ function, denoted by $\sqrt{\sigma_{D^{*+}}}$. 
\begin{align}
    \sqrt{s_\text{b.p.}} = m_{D^{0}} + \sqrt{\sigma_{D^{*+}}}\,,
\end{align}
The singularity has the square-root type in accordance with the $D^*$ zero-width limit.

The effective range expansion as in Eq.~\eqref{eq:scatt.ampl} is not only supposed to approximate the scattering amplitude for physical values of $s$,
but also it needs to match the analytic structure of $\mathcal{T}(s)$.
Hence, the expansion series needs to have a square-root type singularity at $s=s_\text{b.p.}$.
It is achieved by using the break-up momentum $k^*$ with the square-root branch point at the complex plane,
$k^*(s) = \lambda^{1/2}(s,\sigma_{D^{*+}},m_{D^0}^2)/(2\sqrt{s})$.
The regular function $R^*(s)$, analogue to $R(s)$ in Eq.~\eqref{eq:scatt.ampl}, is obtained by removing the singular part from $\mathcal{T}$:
\begin{align} \label{eq:r.matching}
    R^*(s) = N^* \mathcal{T}^{-1}(s) + ik^*(s)\,,
\end{align}
where $N^*$ is a normalization constant.
The Taylor expansion $R^*(s)$ at $s_\text{b.p.}$ proceeds as in Eq.~\eqref{eq:R.NR}:
\begin{align} \label{eq:scatt.range.s*}
    1/a &= R^*(s_\text{b.p.}),\\ \nonumber
    r &= 2 R^{*\prime}(s_\text{b.p.})/k^{2\prime}(s_\text{b.p.})\,,
\end{align}
where the scattering length and effective range are defined using the derivatives with respect to $s$ variable.
The program of the effective-range expansion is a straightforward generalization of the two-body case,
however, there are several challenges to overcome:
\begin{enumerate}
    \item Equation~\eqref{eq:r.matching} required that the direction of the branching cut for the functions $k^*(s)$
    and $N^* \mathcal{T}^{-1}$ are chosen consistently. The direction of the branching cut in the numerical computation of $\mathcal{T}(s)$ is controlled
    by the way how the the path integrals of the the three-body phase-space hit the pole of $D^*$. A dedicated prescription to the path integrals is needed
    \item Computation of the function $R^*$ at the $s_\text{b.p}$ is numerically unstable since the phase-space integral hits the $D^{*+}$ pole exactly. Therefore, a robust numerical method is required for evaluation of Eq.~\eqref{eq:scatt.range.s*} as well as to find the value of $N^*$.
\end{enumerate}


Instead of matching the cut location in $\mathcal{T}(s)$ to the one in $k^*(s)$,
the cut directions are enforced in both functions to a common convention.
The orientation is chosen to be straight down as shown in Fig.~\ref{fig:cuts}.
To achieve it, $k^*(s)$ is computed as a product of four square-root functions,
one of which is ``rotated'' by $90$ degrees by the phase factor, $\exp(-i\pi/2)=-i$, added to the argument of the square-root function.
\begin{align}
    k^*(s) &= \frac{e^{\pi/4}}{2\sqrt{s}}\sqrt{-i\sqrt{s}+i(\sqrt{\sigma_{D^{*+}}}+m_{D^0})} \\ \nonumber
    &\quad \times\sqrt{\sqrt{s}-(\sqrt{\sigma_{D^{*+}}}-m_{D^0})} \\ \nonumber
    &\quad \times\sqrt{\sqrt{s}+(\sqrt{\sigma_{D^{*+}}}+m_{D^0})} \\ \nonumber
    &\quad \times\sqrt{\sqrt{s}+(\sqrt{\sigma_{D^{*+}}}-m_{D^0})}\,.
\end{align}

For evaluation of Eq.~\eqref{eq:rho.i.complex}, we note that the matrix element $\frak{M}^2$ is an analytic
function in $s$, while it contains the $D^*$ poles in $m_{12}$ and $m_{13}$ variables.
The function $\frak{F}(m)$ develops the $D^{*}$ pole in the lower half plane, $\text{Im}\,m < 0$,
while $\frak{F}^\dagger(m)$ includes the pole in the upper half plane, $\text{Im}\,m > 0$~\cite{JPAC:2018zwp}.
To control the border of the phase-space integral which collides with the $D^{*+}$ pole first,
the integral in Eq.~\eqref{eq:rho.i.complex} is split into two parts:
\begin{align} \nonumber
    \int_{\mathcal{D}} dm_{12} dm_{13} &\big[ \frak{F}^\dagger(m_{12}) \frak{X}_{11} \frak{F}(m_{12}) \\\nonumber
    &\quad + \frak{F}^\dagger(m_{13})\,\frak{X}_{21}\,\frak{F}(m_{13}) \\\nonumber
    &\quad + \frak{F}^\dagger(m_{13})\,\frak{X}_{22}\,\frak{F}(m_{13}) \\\nonumber
    &\quad + \frak{F}^\dagger(m_{12})\,\frak{X}_{12}\,\frak{F}(m_{13})\big] \\ \label{eq:frakM.two.parts}
    =\int_{\Gamma_H} dm_{12} \int_{\Gamma_S} dm_{13} &\big[ \frak{F}^\dagger(m_{12}) \,\frak{X}_{11} \,\frak{F}(m_{12}) \\\nonumber
    &\quad + \frak{F}^\dagger(m_{13})\,\frak{X}_{21}\,\frak{F}(m_{12})\big] \\\nonumber
    \,+\int_{\Gamma_H} dm_{13} \int_{\Gamma_S} dm_{12} &\big[\frak{F}^\dagger(m_{13})\,\frak{X}_{22}\,\frak{F}(m_{13}) \\\nonumber
    &\quad + \frak{F}^\dagger(m_{12})\,\frak{X}_{12}\,\frak{F}(m_{13})\big]\,.
\end{align}
where $\Gamma_H$ and $\Gamma_S$ are two special integration paths in the complex $m$ planes shown.
The paths are shown in Fig.~\ref{fig:cuts} for the first term in Eq.~\eqref{eq:frakM.two.parts}.
The ``hook'' path, $\Gamma_H$ for $m_{12}$ variable 
connects $(m_1+m_2)$ and $(\sqrt{s}-m_3)$ with two segments via an intermediate point $\text{Re}(\sqrt{s}-m_3)$.
The ``straight'' path, $\Gamma_S$, in $m_{13}$ variable is a straight line between the Dalitz-plot borders, $m_{13}^+(s,m_{12})$ and $m_{13}^-(s,m_{12})$
presented in Appendix~\ref{sec:integration}.
The expressions are the same the second term in Eq.~\eqref{eq:frakM.two.parts} up to exchange of the indices, $2\leftrightarrow 3$.
The prescription leads to the branch cut in $\sqrt{s}$ plane directed straight down since the $D^{*+}$ pole in $m_{12}$ and $m_{13}$ variables
is always approached from below by the $\Gamma_H$ contour.

Construction of the regular function $R^*$ in Eq.~\eqref{eq:r.matching} requires finding the numerical constant $N^*$
such that the cut contribution cancels on the right side of the equation.
It is done by computing a circular integral around the expansion point, $\sqrt{s_\text{b.p.}}$:
\begin{align}
    N^* = \oint\limits_{|s'-s_\text{b.p.}| = \epsilon}\, \mathcal{T}^{-1}(s')\, ds'\, \bigg/ 
    \oint\limits_{|s'-s_\text{b.p.}| = \epsilon}\,(-ik^*(s'))\, ds'\,.
\end{align}
where $\epsilon$ is a positive radius of the circular integral.
For both integrands, the only singularity enclosed by the contour is a branch cut.
Therefore, both integrals act on the discontinuity of the square-root function which are the same up to the constant $N^*$.
The value of the function $R(s)$ and its derivative at $s=s_\text{b.p.}$ are computed using the Cauchy integral theorem:
\begin{align} \label{eq:cauchy}
R^{*}(s_\text{s.b.})  &= \frac{1}{2\pi i}\oint\limits_{|s'-s_\text{b.p.}|=\epsilon}\, \frac{R^*(s')}{s'-s_\text{b.p.}}\, ds'\,, \\
R^{*\prime}(s_\text{s.b.}) &= \frac{1}{2\pi i}\oint\limits_{|s'-s_\text{b.p.}|=\epsilon}\, \frac{R^*(s')}{(s'-s_\text{b.p.})^2}\, ds'\,.
\end{align}
Equation~\eqref{eq:cauchy} gives a numerically stable procedure and goes around the direct evaluation of the
function $\mathcal{T}(s)$ at the branch point.


According to the LHCb analysis,
the fit prefers large values of the coupling $g$ and does not have sensitivity to its exact value.
The scattering length is computed for \mbox{$g^2\to\infty$} limit and the uncertainty is spanned by the error of the $m_0$ parameter.
The limits of the confidence interval for the effective range are calculated using the interval for $g$: $|g| > 5.1 (4.3)\,$GeV at $90(95)\%\,$CL.
Following the outlined procedure for the effective-range expansion, the values are obtained:
\begin{align}
    1/a &= (-33\pm 2) + (2\pm 0.1)i\,\text{MeV}\,,\\ \nonumber
    r_\infty &< r < r_g^{(\prime)}\text{ at } 90(95)\%\text{ CL}\,.
\end{align}
where $r_\infty$ is the value of the effective range when $g\to \infty$,
and $r_g^{(\prime)}$ corresponds to the low limit value of $g$ at the $90(95)\%$ CL.
\begin{align}
    r_\infty & = -4.3 + 0.5i\,\text{fm}\,,\\\nonumber
    r_g^{(\prime)} & = -16.2 + 0.5i\,(-21.2+0.5i)\,\text{fm}\,.
\end{align}
As anticipated in \cite{LHCb:2021auc},
the scattering length in Eq.~\eqref{eq:a.LHCb}, determined on the real axis,
picks up the imaginary part related to the $D^*$ width rather than the properties of the $T_{cc}^+$ state.
For this reason, the scattering length computed at the complex branch point is closer to the real value.
A shift of the confidence interval of the effective range is encountered with with respect to Eq.~\ref{eq:r.LHCb}. 
The shift value $r_\infty$ is understood as a combination of two effects:
\begin{align} \label{eq:r0}
    r_\infty = r_\text{disp} + r_\text{high}\,. 
\end{align}
where the first term is a dispersion correction and the second is related to the $D^{*0}D^+$ threshold.
These contribution are separated by turning off the coupling of $T_{cc}^+$ to $D^{*0}D^+$ while preserving
the value of the binding energy. 
The found value of $r_\text{high}$ is $-3.7+0.3i\,$fm. It is close to the naive expectation of $-3.78\,$fm obtained with the zero-width approximation for the $D^*$ meson~\cite{Baru:2021ldu}.
The dispersion term $r_\text{disp}$ is related to the left-hand singularities of the break-up momentum function $k(s)$.
In fact, the self-energy function computed using the dispersion technique has only the right-hand cut,
while the non-relativistic expansion is written in term of $k$,
which is singular at the pseudothreshold $\sqrt{s} = (m_{D^{*+}}-m_{D^0})$ and $\sqrt{s}=0$.
The effect is small, $r_\text{disp} = -0.6+0.2i\,$fm since these non-analytic structures are far from the region of interest.

\begin{figure}
    \centering
    \includegraphics[width=\columnwidth]{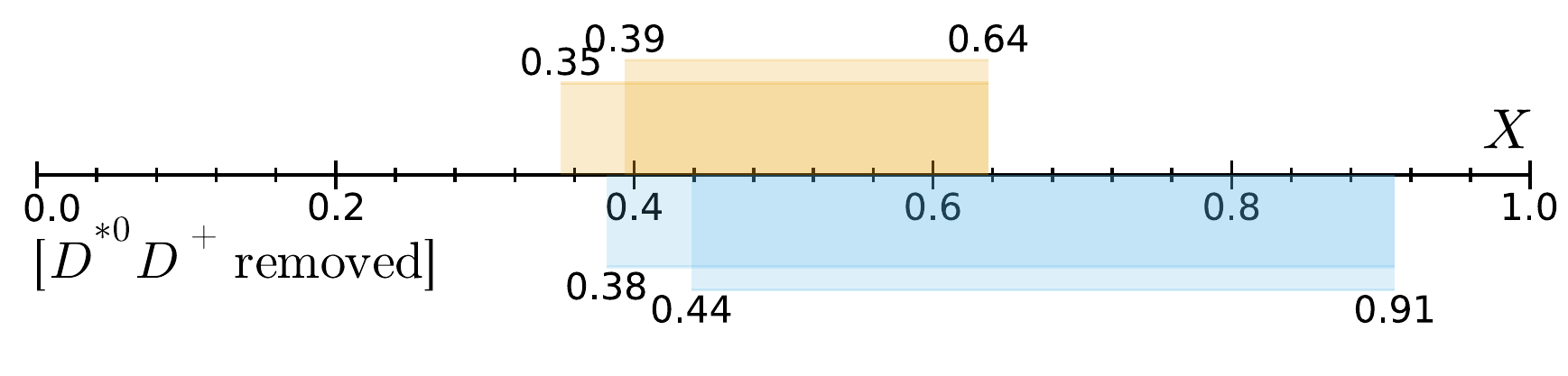}
    \caption{Intervals for the compositeness at $90(95)\,\%$ confidence level.
    The interval based on the the nominal results is shown by shaded orange region above the axis.
    The values of $X$ with excluding $r_\text{high}$ component are shown by the shaded blue regions below the real axis.
    The $D^{*0}D^{+}$ threshold is removed. The dark (light) shaded intervals corresponds to $90(95)\,\%\,$CL for both pictograms.
    }
    \label{fig:compositeness}
\end{figure}

The Weinberg compositeness is evaluated using Eq.~\eqref{eq:compositeness} omitting the small imaginary parts of $r$ and $1/a$.
The confidence intervals for $X$ are shown by shaded orange region in Fig.~\ref{fig:compositeness}.
The blue shaded intervals employ an alternative approach and exclude $r_\text{high}$ component.
Ref.~\cite{Baru:2021ldu} points that the contribution of the higher threshold might need to be removed for the value of the effective range used in the compositeness. The $r_\text{high}$ contribution originates from the isospin-breaking mass difference between $D^{*+}$ and $D^{*0}$.
One finds that the value of $r_\text{high}$ is inverse proportional to the mass difference between the $D^{*+}D^0$ and $D^{*0}D^+$ thresholds and therefore diverges in the isospin limit.
Both approaches lead to rather high values of compositeness of the $T_{cc}^+$ state which is in good agreement with the conclusions of~\cite{LHCb:2021vvq}.
We also note a nice agreement of the realization of the $T_{cc}^+$ hadron in the quark model calculations of Ref.~\cite{Janc:2004qn}.

In summary, the effective-range expansion for the $D^{*}D$ scattering in LHCb model of $T_{cc}^+$ state is studied taking into account three-body nature of the problem. We find that analytic structure of the amplitude is matched by the expansion series only if the break-up momentum  is computed for $D^{*+}D^{0}$ system with the complex mass of $D^{*+}$. In the new expansion, both the scattering length and the effective range have just a small imaginary part. Moreover, the precise procedure reveals two additional contributions to the effective range with respect to the original analysis: a small dispersion correction and relatively large contribution from the $D^{*0}D^+$ threshold.

\begin{acknowledgments}
The project was triggered by a discussion with
Antonio Polosa, Alessandro Pilloni, Angelo Esposito, Luciano Maiani, 
Christoph Hanhart, Alexey Nefediev, Vadim Baru, and Feng-Kun Guo.
The author thanks his LHCb collaborators and especially Ivan Belyaev and Ivan Polaykov for valuable comments to the paper draft.
The work is supported by German Excellence Strategy - EXC-2094 - 390783311.

\end{acknowledgments}

\newpage
\appendix

\section{$T_{cc}^{+}$ amplitude} \label{sec::Tcc.model}
The LHCb model of the $T_{cc}^+$ lineshape is build on the assumption of isoscalar, axial-vector quantum numbers of the state.
Three coupled channels $\pi^+ D^0 D^0$, $\pi^0 D^+ D^0$, $\gamma D^+ D^0$ are considered.
The decays of $T_{cc}^+$ to these states proceed via the $D^*$ resonances in two different particle pairs for all channels.
In the model, couplings of $T_{cc}^+$ to $D^{*+} D^0$ and $D^{*0}D^+$ are equal by the absolute value and opposite by the sign.
The squared $T_{cc}^+$ decay matrix element is cast to a bilinear form,
\begin{align}
    \frak{M}_f^2(s,m_{12}^2,m_{13}^2) = \frak{F}_f^\dagger\,\frak{X}_f(s,m_{12}^2,m_{13}^2)\,\frak{F}_f\,,
\end{align}
where $\frak{F}$ is a two-element vector of $D^*$ amplitudes,
$\frak{X}(s,m_{12}^2,m_{13}^2)$ is a two dimensional squared matrix of functions polynomial in mandelstam variables,
$m_{12}^2$ and $m_{13}^2$ and $s$. Specifically,
\begin{align} \nonumber
    \frak{F}_{\pi^+ D^0 D^0} &=
    \begin{pmatrix}
    \frak{F}^{+}(m_{12}^2)\\
    \frak{F}^{+}(m_{13}^2)
    \end{pmatrix}\,,
&    %
    \frak{F}_{\pi^0(\gamma) D^+ D^0} &=
    \begin{pmatrix}
    \frak{F}^{+}(m_{12}^2)\\
    \frak{F}^{0}(m_{13}^2)
    \end{pmatrix}\,,
\end{align}
where $D^*$ propagators are paramertized by the relativistic Breit-Wigner amplitude:
\begin{align}
    \frak{F}^{c}(m^2) = \frac{1}{m_{D^{*c}}^2-m^2 - i m_{D^{*c}} \Gamma_{D^{*c}}}\,,
\end{align}
and $c$ indicates the change, $c=+,0$. The PDG values~\cite{ParticleDataGroup:2020ssz} are used masses and widths of $D^{*+}$ and $D^{*0}$.

The symmetric squared matrices $\frak{X}$ are expressed via the invariant functions, $A$, $B$, $C$, and $G$ given in~\cite{LHCb:2021vvq} (see Eq.~(M7)).
The form of the elements of the $\frak{X}$ matrix are the same for $\pi^+ D^0 D^0$ and $\pi^0 D^+ D^0$ system.
\begin{align}
    \left( \frak{X}_{\pi^+ D^0 D^0} \right)_{11} = \left( \frak{X}_{\pi^0 D^+ D^0} \right)_{11} &= \frac{f^2}{12} A\,,\\ \nonumber
    \left( \frak{X}_{\pi^+ D^0 D^0} \right)_{22} = \left( \frak{X}_{\pi^0 D^+ D^0} \right)_{22} &= \frac{f^2}{12} B\,,\\ \nonumber
    \left( \frak{X}_{\pi^+ D^0 D^0} \right)_{12} = \left( \frak{X}_{\pi^0 D^+ D^0} \right)_{12} &= \frac{f^2}{24} C\,.
\end{align}
For $\gamma D^0 D^0$, it reads:
\begin{align}\nonumber
    &\left( \frak{X}_{\gamma D^0 D^0} \right)_{11} = \frac{\mu_+^2 h^2}{3} \left(\frac{(m_{12}^2-m_1^2-m_2^2)^2}{4} + G \right)\,,\\ \nonumber
    &\left( \frak{X}_{\gamma D^0 D^0} \right)_{22} = \frac{\mu_0^2 h^2}{3} \left(\frac{(m_{13}^2-m_1^2-m_3^2)^2}{4}+ G\right)\,,\\ 
    &\left( \frak{X}_{\gamma D^0 D^0} \right)_{12} = \frac{\mu_+\mu_0 h^2}{3} \\ \nonumber
    &\qquad\times\left(
    \frac{
    (m_{12}^2-m_1^2-m_2^2)
    (m_{13}^2-m_1^2-m_3^2)}{4} - G\right)\,.
\end{align}
The $D^*$ decay constants $f^2 = 282.42$ and $h^2 = 20.13\times 10^{-3}\,\text{MeV}^{-2}$ are derived from the widths of $D^{*+}$ and $D^{*0}$ mesons, respectively~\cite{LHCb:2021vvq,ParticleDataGroup:2020ssz}. The magnetic moments are fixed to be, $\mu_+ = 1$, $\mu_0 = -3.77$ according to~\cite{Rosner_2013, Gasiorowicz:1981jz, Rosner:1980bd}.

\section{Integration contours} \label{sec:integration}
The integration of the three-body phase space for complex value of the total energy is carried out in $m_{12}$ and $m_{13}$ variables.
For the $m_{12}$ integral, the integration path connects the points $(m_1+m_2)^2$ and $(\sqrt{s}-m_3)$, while
the ranges of the $m_{13}$ integral are determined by the Dalitz plot borders $m_{13}^{\pm}(s,m_{12})$. An explicit factorization of the squared-root argument gives an exact prescription of the branch cut location in the $m_{13}$ variable.
\begin{align} \label{eq:m13pm}
    &m_{13}^{2\pm}(s,m_{12}) = m_1^2+m_3^2 \\\nonumber
    &\quad +\frac{(m_{12}^2+m_1^2-m_2^2)(s-m_{12}^2-m_3^2)}{2m_{12}^2} \\\nonumber
    &\quad \pm\frac{1}{2m_{12}^2}\sqrt{m_{12}-(m_1+m_2)}\sqrt{m_{12}-(m_1-m_2)}\\\nonumber
    &\quad\quad \times \sqrt{m_{12}+(m_1+m_2)}\sqrt{m_{12}+(m_1-m_2)}\\\nonumber
    &\quad\quad \times \sqrt{m_{12}-(\sqrt{s}-m_3)}\sqrt{m_{12}-(\sqrt{s}+m_3)}\\\nonumber
    &\quad\quad \times \sqrt{m_{12}+(\sqrt{s}-m_3)}\sqrt{m_{12}+(\sqrt{s}+m_3)}\,.
\end{align}

\bibliographystyle{h-physrev}
\bibliography{ref}

\begin{thebibliography}{10}

\bibitem{ParticleDataGroup:2020ssz}
Particle Data Group, P.~A. Zyla {\em et~al.},
\newblock PTEP {\bf 2020}, 083C01 (2020).

\bibitem{Olsen:2017bmm}
S.~L. Olsen, T.~Skwarnicki, and D.~Zieminska,
\newblock Rev. Mod. Phys. {\bf 90}, 015003 (2018), 1708.04012.

\bibitem{Brambilla:2019esw}
N.~Brambilla {\em et~al.},
\newblock Phys. Rept. {\bf 873}, 1 (2020), 1907.07583.

\bibitem{Guo:2017jvc}
F.-K. Guo {\em et~al.},
\newblock Rev. Mod. Phys. {\bf 90}, 015004 (2018), 1705.00141.

\bibitem{Esposito:2016noz}
A.~Esposito, A.~Pilloni, and A.~D. Polosa,
\newblock Phys. Rept. {\bf 668}, 1 (2017), 1611.07920.

\bibitem{Ali:2019roi}
A.~Ali, L.~Maiani, and A.~D. Polosa,
\newblock {\em {Multiquark Hadrons}} (Cambridge University Press, 2019).

\bibitem{Chen:2016qju}
H.-X. Chen, W.~Chen, X.~Liu, and S.-L. Zhu,
\newblock Phys. Rept. {\bf 639}, 1 (2016), 1601.02092.

\bibitem{Ali:2017jda}
A.~Ali, J.~S. Lange, and S.~Stone,
\newblock Prog. Part. Nucl. Phys. {\bf 97}, 123 (2017), 1706.00610.

\bibitem{Liu:2019zoy}
Y.-R. Liu, H.-X. Chen, W.~Chen, X.~Liu, and S.-L. Zhu,
\newblock Prog. Part. Nucl. Phys. {\bf 107}, 237 (2019), 1903.11976.

\bibitem{Belle:2003nnu}
Belle, S.~K. Choi {\em et~al.},
\newblock Phys. Rev. Lett. {\bf 91}, 262001 (2003), hep-ex/0309032.

\bibitem{LHCb:2021auc}
LHCb, R.~Aaij {\em et~al.},
\newblock (2021), 2109.01056.

\bibitem{LHCb:2021vvq}
LHCb, R.~Aaij {\em et~al.},
\newblock (2021), 2109.01038.

\bibitem{SilvestreBrac:1993ss}
B.~Silvestre-Brac and C.~Semay,
\newblock Z. Phys. C {\bf 57}, 273 (1993).

\bibitem{Pepin:1996id}
S.~Pepin, F.~Stancu, M.~Genovese, and J.~M. Richard,
\newblock Phys. Lett. B {\bf 393}, 119 (1997), hep-ph/9609348.

\bibitem{Lee:2009rt}
S.~H. Lee and S.~Yasui,
\newblock Eur. Phys. J. C {\bf 64}, 283 (2009), 0901.2977.

\bibitem{Janc:2004qn}
D.~Janc and M.~Rosina,
\newblock Few Body Syst. {\bf 35}, 175 (2004), hep-ph/0405208.

\bibitem{Ebert:2007rn}
D.~Ebert, R.~N. Faustov, V.~O. Galkin, and W.~Lucha,
\newblock Phys. Rev. D {\bf 76}, 114015 (2007), 0706.3853.

\bibitem{Vijande:2003ki}
J.~Vijande, F.~Fernandez, A.~Valcarce, and B.~Silvestre-Brac,
\newblock Eur. Phys. J. A {\bf 19}, 383 (2004), hep-ph/0310007.

\bibitem{Luo:2017eub}
S.-Q. Luo, K.~Chen, X.~Liu, Y.-R. Liu, and S.-L. Zhu,
\newblock Eur. Phys. J. C {\bf 77}, 709 (2017), 1707.01180.

\bibitem{Park:2018wjk}
W.~Park, S.~Noh, and S.~H. Lee,
\newblock Nucl. Phys. A {\bf 983}, 1 (2019), 1809.05257.

\bibitem{Deng:2018kly}
C.~Deng, H.~Chen, and J.~Ping,
\newblock Eur. Phys. J. A {\bf 56}, 9 (2020), 1811.06462.

\bibitem{Yang:2019itm}
G.~Yang, J.~Ping, and J.~Segovia,
\newblock Phys. Rev. D {\bf 101}, 014001 (2020), 1911.00215.

\bibitem{Maiani:2019lpu}
L.~Maiani, A.~D. Polosa, and V.~Riquer,
\newblock Phys. Rev. D {\bf 100}, 074002 (2019), 1908.03244.

\bibitem{Tan:2020ldi}
Y.~Tan, W.~Lu, and J.~Ping,
\newblock Eur. Phys. J. Plus {\bf 135}, 716 (2020), 2004.02106.

\bibitem{Lu:2020rog}
Q.-F. L\"u, D.-Y. Chen, and Y.-B. Dong,
\newblock Phys. Rev. D {\bf 102}, 034012 (2020), 2006.08087.

\bibitem{Faustov:2021hjs}
R.~N. Faustov, V.~O. Galkin, and E.~M. Savchenko,
\newblock Universe {\bf 7}, 94 (2021), 2103.01763.

\bibitem{Noh:2021lqs}
S.~Noh, W.~Park, and S.~H. Lee,
\newblock Phys. Rev. D {\bf 103}, 114009 (2021), 2102.09614.

\bibitem{Feng:2013kea}
G.~Q. Feng, X.~H. Guo, and B.~S. Zou,
\newblock (2013), 1309.7813.

\bibitem{Navarra:2007yw}
F.~S. Navarra, M.~Nielsen, and S.~H. Lee,
\newblock Phys. Lett. B {\bf 649}, 166 (2007), hep-ph/0703071.

\bibitem{Wang:2017uld}
Z.-G. Wang,
\newblock Acta Phys. Polon. B {\bf 49}, 1781 (2018), 1708.04545.

\bibitem{Gao:2020ogo}
D.~Gao {\em et~al.},
\newblock (2020), 2007.15213.

\bibitem{Semay:1994ht}
C.~Semay and B.~Silvestre-Brac,
\newblock Z. Phys. C {\bf 61}, 271 (1994).

\bibitem{Carlson:1987hh}
J.~Carlson, L.~Heller, and J.~A. Tjon,
\newblock Phys. Rev. D {\bf 37}, 744 (1988).

\bibitem{Li:2012ss}
N.~Li, Z.-F. Sun, X.~Liu, and S.-L. Zhu,
\newblock Phys. Rev. D {\bf 88}, 114008 (2013), 1211.5007.

\bibitem{Liu:2019stu}
M.-Z. Liu, T.-W. Wu, M.~Pavon~Valderrama, J.-J. Xie, and L.-S. Geng,
\newblock Phys. Rev. D {\bf 99}, 094018 (2019), 1902.03044.

\bibitem{Cheng:2020wxa}
J.-B. Cheng, S.-Y. Li, Y.-R. Liu, Z.-G. Si, and T.~Yao,
\newblock Chin. Phys. C {\bf 45}, 043102 (2021), 2008.00737.

\bibitem{Karliner:2017qjm}
M.~Karliner and J.~L. Rosner,
\newblock Phys. Rev. Lett. {\bf 119}, 202001 (2017), 1707.07666.

\bibitem{Braaten:2020nwp}
E.~Braaten, L.-P. He, and A.~Mohapatra,
\newblock Phys. Rev. D {\bf 103}, 016001 (2021), 2006.08650.

\bibitem{Baru:2003qq}
V.~Baru, J.~Haidenbauer, C.~Hanhart, Y.~Kalashnikova, and A.~E. Kudryavtsev,
\newblock Phys. Lett. B {\bf 586}, 53 (2004), hep-ph/0308129.

\bibitem{Weinberg:1962hj}
S.~Weinberg,
\newblock Phys. Rev. {\bf 130}, 776 (1963).

\bibitem{Weinberg:1965zz}
S.~Weinberg,
\newblock Phys. Rev. {\bf 137}, B672 (1965).

\bibitem{Bethe:1949yr}
H.~A. Bethe,
\newblock Phys. Rev. {\bf 76}, 38 (1949).

\bibitem{Kallen}
G.~K{\"{a}}ll{\'{e}}n,
\newblock {\em {Elementary particle physics}} (Addison\nobreakdash-Wesley,
  Reading, Massachusetts, 1964).

\bibitem{Smorodinsky:1948xyz}
Y.~A. Smorodinsky,
\newblock Dokl. Akad. Nauk SSSR {\bf 60}, 217 (1948).

\bibitem{Landau:1991wop}
L.~D. Landau and E.~M. Lifshits,
\newblock {\em {Quantum Mechanics}: {Non-Relativistic Theory}}, Course of
  Theoretical Physics Vol. v.3 (Butterworth-Heinemann, Oxford, 1991).

\bibitem{Wigner:1955zz}
E.~P. Wigner,
\newblock Phys. Rev. {\bf 98}, 145 (1955).

\bibitem{Esposito:2021vhu}
A.~Esposito, L.~Maiani, A.~Pilloni, A.~D. Polosa, and V.~Riquer,
\newblock (2021), 2108.11413.

\bibitem{Bruns:2019xgo}
P.~C. Bruns,
\newblock (2019), 1905.09196.

\bibitem{Baru:2021ldu}
V.~Baru {\em et~al.},
\newblock (2021), 2110.07484.

\bibitem{Sekihara:2014kya}
T.~Sekihara, T.~Hyodo, and D.~Jido,
\newblock PTEP {\bf 2015}, 063D04 (2015), 1411.2308.

\bibitem{Matuschek:2020gqe}
I.~Matuschek, V.~Baru, F.-K. Guo, and C.~Hanhart,
\newblock Eur. Phys. J. A {\bf 57}, 101 (2021), 2007.05329.

\bibitem{Li:2021cue}
Y.~Li, F.-K. Guo, J.-Y. Pang, and J.-J. Wu,
\newblock (2021), 2110.02766.

\bibitem{Song:2022yvz}
J.~Song, L.~R. Dai, and E.~Oset,
\newblock (2022), 2201.04414.

\bibitem{LHCb:2020xds}
LHCb, R.~Aaij {\em et~al.},
\newblock Phys. Rev. D {\bf 102}, 092005 (2020), 2005.13419.

\bibitem{Hanhart:2007yq}
C.~Hanhart, Y.~S. Kalashnikova, A.~E. Kudryavtsev, and A.~V. Nefediev,
\newblock Phys. Rev. D {\bf 76}, 034007 (2007), 0704.0605.

\bibitem{Albaladejo:2021vln}
M.~Albaladejo,
\newblock (2021), 2110.02944.

\bibitem{Du:2021zzh}
M.-L. Du {\em et~al.},
\newblock Phys. Rev. D {\bf 105}, 014024 (2022), 2110.13765.

\bibitem{meeting.CERN}
Effective range for $x(3872)$,
\newblock \url{https://indico.cern.ch/event/1086301/},
\newblock A topical workshop, CERN, Switzerland.

\bibitem{meeting.Munich}
Hadronic three-body christmas,
\newblock \url{https://indico.ph.tum.de/event/6909/},
\newblock A topical workshop, Munich, Germany.

\bibitem{Mikhasenko:2019vhk}
M.~Mikhasenko {\em et~al.},
\newblock JHEP {\bf 08}, 080 (2019), 1904.11894.

\bibitem{JPAC:2018zwp}
JPAC, M.~Mikhasenko {\em et~al.},
\newblock Phys. Rev. D {\bf 98}, 096021 (2018), 1810.00016.

\bibitem{Rosner_2013}
J.~L. Rosner,
\newblock Phys. Rev. {\bf D88}, 034034 (2013), 1307.2550.

\bibitem{Gasiorowicz:1981jz}
S.~Gasiorowicz and J.~L. Rosner,
\newblock Am. J. Phys. {\bf 49}, 954 (1981).

\bibitem{Rosner:1980bd}
J.~L. Rosner,
\newblock NATO Sci. Ser. B {\bf 66}, 1 (1981).

\end{thebibliography}

\end{document}